\documentclass[intlimits,twoside,a4paper]{article}

\usepackage{graphicx,psfrag}

\usepackage[T2A]{fontenc}
\usepackage[cp1251]{inputenc} 

\usepackage[eqsecnum]{cmpj3}

\RequirePackage{color}

\issue{2024}{27}{3}{33101}
\doinumber{10.5488/CMP.27.33101}

\title[Richard Kirwan]%
{Richard Kirwan a [united] Irish man of science in Europe}
\author[R. Folk]{R. Folk\orcid{0000-0002-5958-4106}\thanks{Email: \email{r.folk@liwest.at}.}}
        \address{ Institute for Theoretical Physics, Johannes Kepler University Linz, 4040 Linz, Austria}

\date{Received November 21, 2023, in final form January 02, 2024}

\begin{document}
\maketitle
\begin{abstract}
The late eighteenth and early nineteenth centuries have long been considered as a formative period for modern Irish political traditions such as nationalism, republicanism and unionism. For Europe it was the time of a turnover in science moving from observation to experiment and from speculation to fact. Richard Kirwan was a well known natural philosopher in Europe and a respected man of science in his time. Throughout all the wars, he was connected with his colleagues in a network reaching across Europe and even to America. Using a few examples, this article is intended to provide an insight how the network worked in a time that was marked by political conflicts and revolutionary events in both science and social life.

\end{abstract}

\section{A late appearance in the world of scientific publications \label{late}}

\begin{quote}{\sc It may be said, with equal truth, of the progress through science, as of that through space, that it is impossible to arrive at any distant point without passing through the intermediate, however great we may suppose the energy of motion in the one case, or that of genius in the other.} (Richard Kirwan, 1780~\cite{Scheele1780})
\end{quote}

It was only at the age of 47 that Richard Kirwan (1733--1812) made his first contribution to the world of publications. He was invited to include Notes to the English translation of Carl Wilhelm Scheele's (1742--1786) book from 1777 {\it Chemische Abhandlung von der Luft und dem Feuer (Chemical Observations and Experiments on Air and Fire)} together with a prefatory introduction by Tobern Bergman (1735--1784)~\cite{Scheele1780}. This invitation may have come at an initiative\footnote{In the letter to Kirwan, included in the book, Priestley wrote: I am far from pretending to a complete knowledge of chemistry, but Dr. Forster's translation, and your Notes together, seem to have answered all my wishes.} of Joseph Priestley's (1733--1804).   The translator John-Reinold Forster in his dedication to Priestley gives further background details how the translation came about: ``No sooner was the present Treatise of Mr. Scheele, about two years ago, transmitted to England from Germany ... which has in many respects so great a reference to your  immortal discoveries on Air [Priestley found in 1774 that the element Air is a mixture of gases containing as one part oxygen (dephlogisticated air)]. The norther philosopher [Scheele] has treated the subject as a Chemist, you as a Philosopher; ... you knew nothing of his Experiments, and he was ignorant of your great and numerous discoveries, when he made his Experiments... The differences in our Author's performance arise chiefly from his not attending always to some principles of natural Philosophy, or else from adhering too scrupulously to chemistry, and not so much founding certain inferences on Experiment as on reasoning.''

\begin{figure} 
\centering\includegraphics[height=7.5cm]{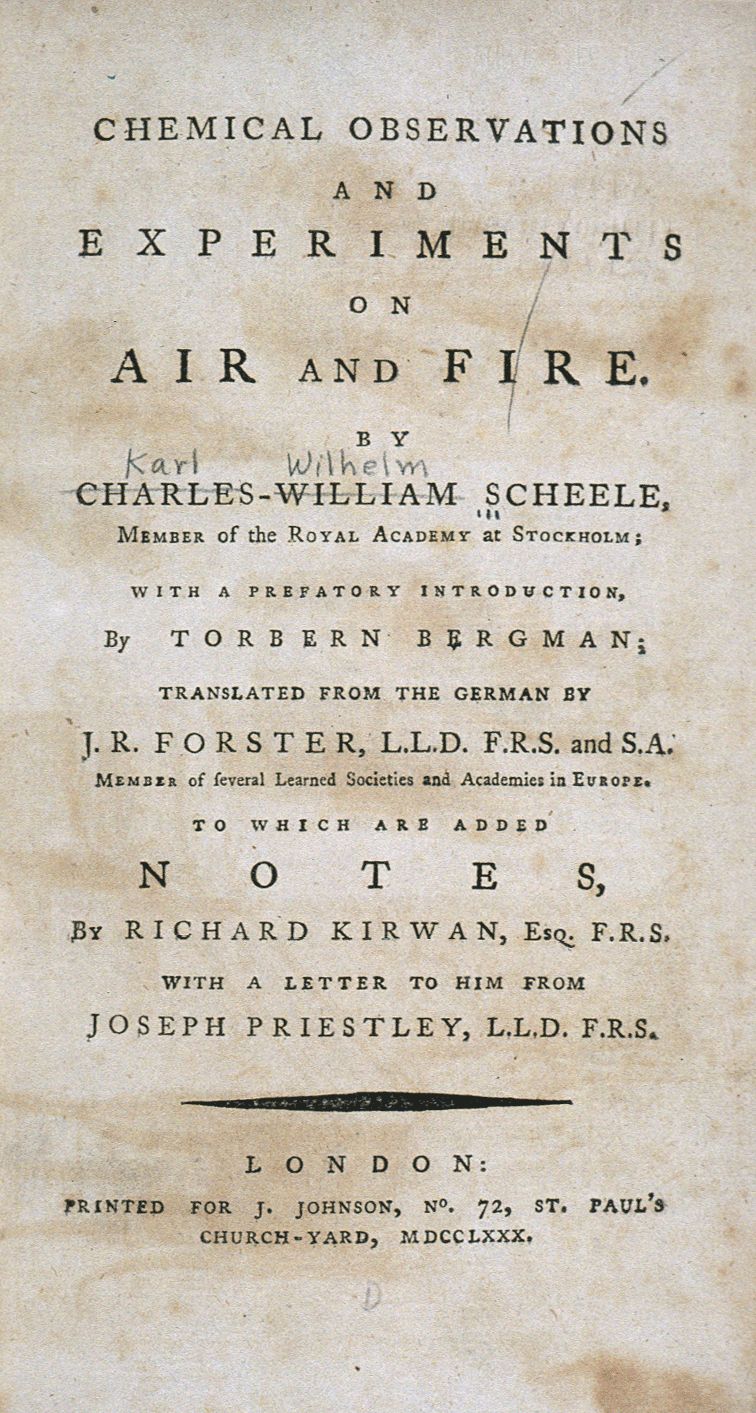}\hspace{0.05cm}\includegraphics[height=7.5cm]{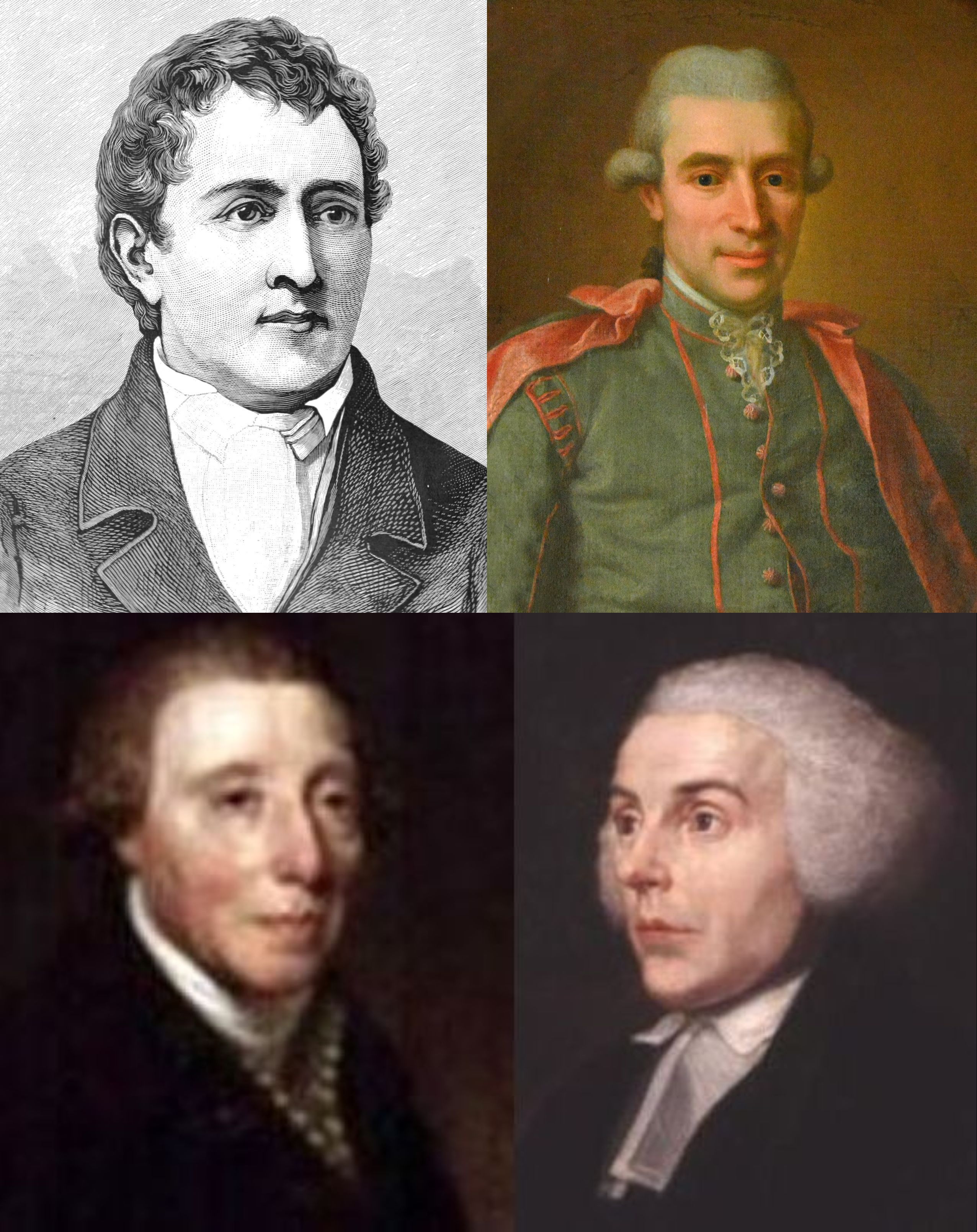}
\caption{(Colour online) (a) English translation from 1780 of Scheele's {\it Abhandlung von Luft und Feuer} from 1777 and the contributers (b) Scheele, Bergman, Kirwan, Priestley. \label{scheele}}
\end{figure}

The book was so topical, that a French edition was published the following year\footnote{The British chemist T. E. Thorpe wrote in 1892 according to the 150th anniversary of Scheele's birth: Had Scheele possessed that sense of quantitative accuracy which was the special characteristic of his contemporary Cavendish, his work on {\it Air and Fire} would inevitably have effected the overthrow of phlogistonism long before the advent of Lavoisier.}. It was translated by Baron Philippe Fr\'ederich de Dietrich (1748--1793) and contained in addition a paper by Scheele from 1779. A second German edition (figure~\ref{scheele}) was published by Crusius in 1782 including the contributions of the English edition, an essay of Gottfried Leonhardi, and the paper by Scheele (for more information on Scheele and Bergman see  \cite{Lennartson2020}).

This late appearance of Kirwan is surprising but he was already well known as scientist\footnote{In the same year he became Fellow of the Royal Society of London.}. And not only his scientific curriculum shows what he mentioned at the beginning of his contribution and was cited above: ``many instances of this sort [capital discoveries based on the former work] occur in the History of the Sciences, nor is there anything that ought to encourage us more in our pursuits of natural knowledge, than the chance that perhaps one step further may be a capital discovery, and the certainty, that if it be not, it at least necessarily leads towards one.''  And phlogiston\footnote{Phlogiston was already introduced in 1667 by Johann Joachim Becher (1635--1682) as a substance which is {\bf set free} in the combustion process. Since Lavoisier and the new chemistry we know instead oxygen is {\bf added} in this process. } theory was one of those steps between alchemy and the modern chemistry of the 19th century.

In the times of Richard Kirwan's life, from the second half of the eighteenth century (when he had finished his university education in 1755) to the beginning of the nineteenth century (when he died in 1812) there was a turnover in natural sciences to
scientific disciplines took place. It was indeed the age when knowledge was in ferment,  said  G.~S.~Rousseau and Roy Poter \cite{ferment}.  Other historians saw it  {\it a century not of peaks but of trackless bog}, so to say `nothing' between Newton and Maxwell. But  it is  important as Kirwan said, to keep going on the process of research.

In his introduction to Scheel's book, Bergman describes three degrees of the Science of Nature (1) the {\sc outsides} of the natural bodies [the alphabet of the great book of nature] are the objects of {\sc Natural History}, (2) the {\sc general qualities} of matter [the spelling] are treated by {\sc Natural Philosophy} (Physica) and (3) the {\sc innermost part} --- the material elements --- [the reading distinctly] are examined by {\sc Chemistry}. Kirwan mentioned the Book of Nature in 1789 in the English new translation \cite{kirwanPhlogiston1789} of the French edition of his essay: 

\begin{quote}{\sc In my opinion the book of nature should be interpreted like other books, the sense of which must be collected not from single detached pages, but from an attentive consideration of the whole, ...}\end{quote}

This sounds different from Galilei's opinion. He said \cite{galilei1623}: {\it ``The grand book of the universe ... cannot be understood unless one first learns to comprehend the language and to read the alphabet in which it is composed ... the language of mathematics.''} And this suspicion against mathematics is also found in the constitution of the Chapter Coffee House Philosophical Society\footnote{In December 1780 a group of intellectuals under the leadership of Richard Kirwan decided to meet fortnightly for discussing Natural Philosophy.} which does not allow consideration of subjects that may lead to mathematical disquisitions rather being confined to Experimental Philosophy, e.g., topics concerning astronomy should not be considered \cite{levere2002}.

\section{Who was Richard Kirwan?}

\begin{quote}{\sc How great must have been the pleasure of a Napier,  a Briggs, a Newton and a Bernoui[!]lli, while intent on the  most laborious calculations? or of a Boyle, a Black, a  Priestly, and a Lavoisier, discovering and scrutinizing the  invisible agents of nature, as Newton did the connecting  principle of the stupendous masses that surround us.} (Richard Kirwan 1810 \cite{kirwan1810})\end{quote}

Richard Kirwan was born in 1733 at  Cloughballymore (south of Galway), Co.~Galway in the far west of Ireland, as second of four sons of a prominent wealthy family. Already in his early years he read chemistry books and tried out his own experiments. He was first educated at home and then later went to the Erasmus Smith school (Galway Grammar School).
In order to proceed his academic studies he had to leave Ireland due to his Catholic Faith\footnote{Irish Colleges were set up to educate Roman Catholics from Ireland in their own religion following the takeover of the country by the Protestant English state in the Tudor conquest of Ireland. In Poitiers it was founded in 1674 and closed in 1762. Even before the official establishment Irish clerical students would have studied at the University of Poitiers.} and the Penal  Laws\footnote{Due to this law Irish Catholics were not allowed to study at  Universities of the Irish or British Kingdom  This is only one aspect of the colonialism. {\it In Ireland, therefore, the Gaelic language, Gaelic customs, Gaelic mythologies, Brehon laws, Gaelic social structures and Catholic affiliation were all continually and successfully undermined through ongoing colonial plantation, disenfranchisement, confiscation and the Penal Law suppressions }\cite{Kenna23}.} \cite{penaltyLaw1,penaltyLaw2}. He followed  in 1750 his brother Patrick and went first to the University of Poitiers but later from 1754 to 1755 he continued his studies in Paris. There he visited the lectures\footnote{Kirwan's Notes taken at these lectures are held at the Royal Irish Academy in Dublin \cite{KnotesR1754}.} of Guillaume-Francois Rouelle (1703--1770), who taught a whole generation of French chemists including Lavoisier \cite{gkim}. 

When his brother was killed [some say it was in a duel] Richard returned to Ireland to take over the family estate Cregg Castle at Corrandulla [Cor an Dola]  (north of Galway). He continued his chemical studies and made some new discoveries himself.  He tried to contact Joseph Black (1728--1799) to exchange with him his results but Black did not answer\footnote{Black of Irish patronage was  appointed professor at the University of Glasgow in 1757. He found in 1756 carbon dioxide {\sc fixed air} as part of the `Element Air', about 1762 latent heat when water changed its phase and the specific heat of substances. But he never published his results.}. This led to such a disappointment that Richard cancelled his chemical experiments in the following time. Anyway later Kirwan and Black became friends.

In 1757 Richard married Anne Blake from Menlough Castle with whom he had two daughters. He lived with her at Blake's estate Menlough. There he continued his studies and enlarged his library. That year, during a visit to Paris, he developed a fever and after the visit he suffered from difficulty swallowing. This caused him great discomfort for most of the rest of his life, as is also reported in a letter from Schwediauer\footnote{Born in Steyr, Upper Austria, as son of a Swedish family. He qualified in Vienna University as physician. His ties to British science began in the 1770s, when he travelled to England and visited London, Edinburgh, and the Lunar Society before finally in 1789 settled in Paris and became French. He came in contact with Kirwan in discussions on some translation project, which were not realized.} (1748--1824) to Tobern Bergman from 1782 \cite{linder1968}. 

He decided to change his situation in 1766 and to study Law in England and Germany. Unfortunately there is no information where he stayed in Europe in this time. It is reported that he traveled to Paris for his wife's health. In order to come along with the conditions\footnote{Again the penal laws did not allow an Irish Catholic to become a barrister.}  to become a barrister, he had to convert. After a period of questioning and and hesitation he abjured his faith in 1764  and became an Anglican. 

Thus, he could proceed with his law studies, but his wife died unexpectedly in 1765 while he was in London. Without knowledge about her illness he only returned two weeks after her funeral --- a situation by which he was probably deeply afflicted. Finally he was called in 1766  to the Irish Bar. But he practiced only two years and took up his scientific research again.  The years from 1769 to 1772 he again spent on the continent, mainly in Paris, but maybe also in Germany. When he returned to Ireland he lodged in Dublin, Peter Street to be near the French Convent where his daughters were educated.

\section{The Year 1777}  

\subsection{The battle of Saratoga}

 \begin{quote}{\sc \dots the sciences are never at war \dots}
(Antoine Laurent Lavoisier to Joseph Lakanal,~1793~\cite{Lavoisier1793})
\end{quote}

Elise Lipkowitz demonstrates in her thesis \cite{Lipkowitz2009} that the events, institutional changes, and political battles of the war years shifted several practices central to scientific life: the geography and politics of letter writing, the ownership and status of scientific objects, as well as the movement and activities of savants. There was already the ongoing Revolutionary War (1775--1783) against Britain establishing and securing the independence of the United States. 

The battle of Saratoga on October 7th 1777 was a key victory for the Americans in the Revolutionary War. Brittain's General John Burgoyne had to surrender his army and {\it this was a great turning point of the war because it won for Americans the foreign assistance which was the last element needed for victory} \cite{morgan1956}. 1776 Benjamin Franklin (1706--1790) was sent to Paris by the Congress to negotiate a treaty of alliance with France. But independent of this political important mission he could communicate with his French scientific colleagues. So he was present in the experiments which Lavoisier reported  to the Academy in September, 1777 because he was invited by him.  Lavoisier showed on certain substances that under normal atmospheric conditions they are in a gaseous state and he concluded\footnote{It would be too simplistic to interpret this as phase transitions as we understand them today. The early history of the phase diagram of a substance began later in 1822 with Cagniard de la Tour \cite{BHK2009}.} {\it all substances in nature can exist in three different states: as a solid, as a liquid and as an `aerial fluid' (fluide a\'eriforme). The degree of heat alone is sufficient to convert the same substance into one
of these three states in succession.}

\begin{figure}
\includegraphics[height=5.0cm]{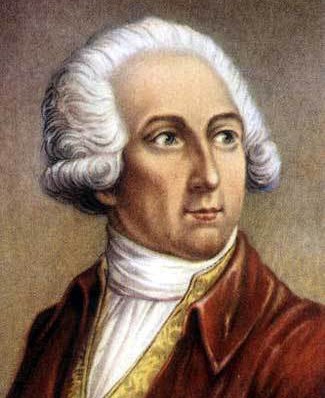}\hspace{0.05cm}\includegraphics[height=5.0cm]{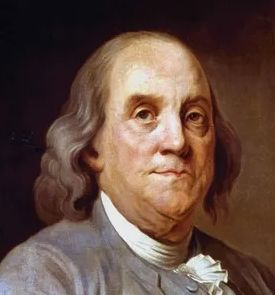}\centering\includegraphics[height=5.0cm]{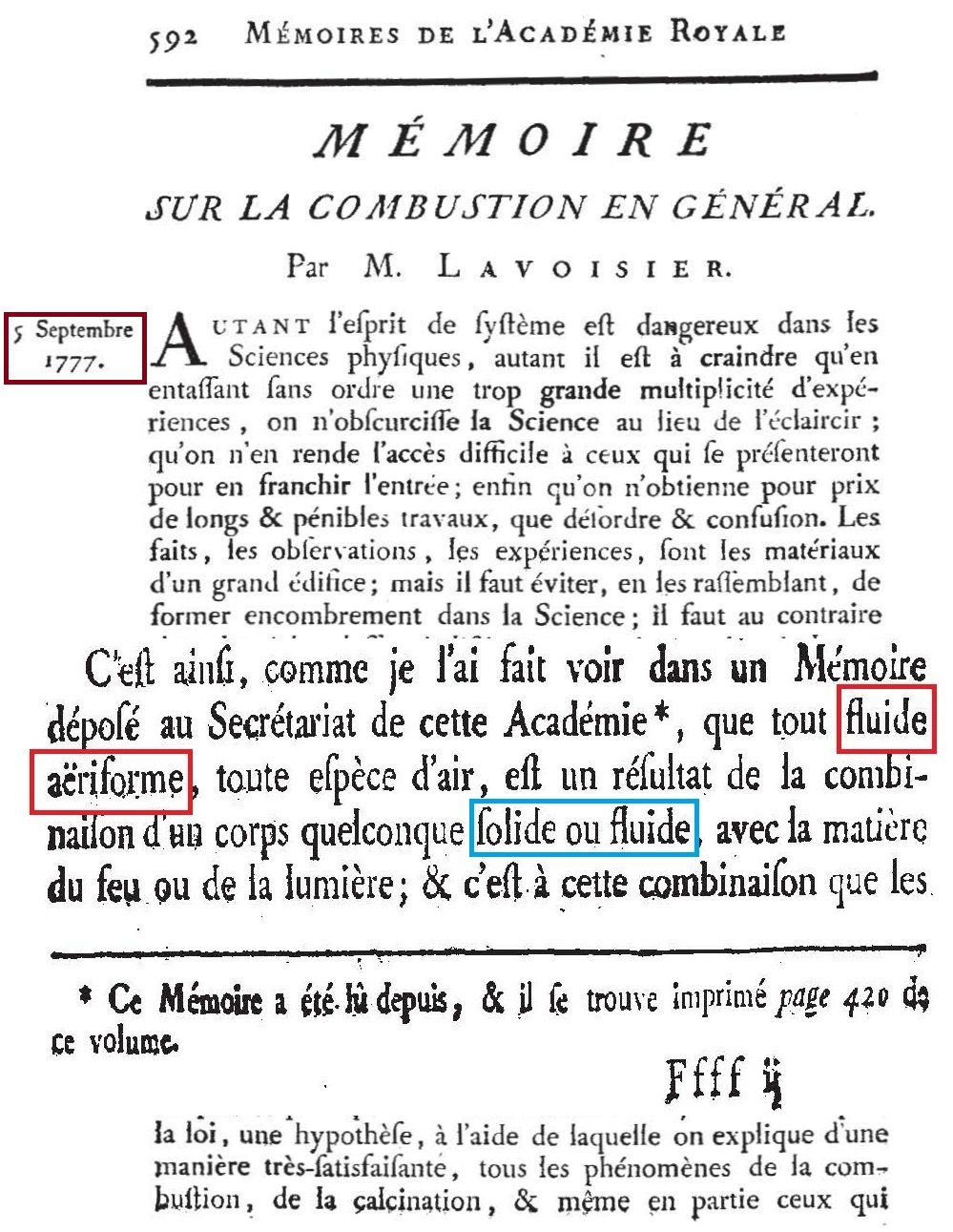}
\caption{(Colour online) Lavoisier invited Benjamin Franklin to be present when he repeated some experiments on air of Mr. Priestley. These experiments were reported to the Academy and published \cite{lavoisier1777}.}
\end{figure}

On 4 December 1777, Benjamin Franklin received two pieces of news in Versailles, on the one hand that Philadelphia had been taken by the British, and on the other hand about the victory at the Battle of Saratoga. Two days later, Louis XVI assented to negotiations for an alliance. The treaty was signed on 6 February 1778, and France declared war on Britain one month later. George III did not welcome a war with France, but he was ``prepared'' for it remembering British victories over that Bourbon power in the Seven Years' War~\cite{Wikipedia}.

Benjamin Franklin, however, used his stay in Europe to visit his friend Joseph Priestley with whom he was in contact also as a member of the Lunar Society of Birmingham (short Lunar Men\footnote{The name comes from the meetings when there is full moon in order to make travelling at night more secure.}) where Richard Kirwan also was member.

Maurice Crosland \cite{Crosland2005} characterized the situation:
{\it  A major problem in communication and cooperation throughout the eighteenth century was the outbreak of intermittent wars. Britain and France were now the two leading  world powers. The constant rivalry between these two great powers often led to war. It  has been calculated that half the period from 1689 to 1815, which we might call the long  eighteenth century, was marked by war between the two neighbouring countries.}  Quite recently on Edward N. Luttwak \cite{luttwak2022} gave a much more pointed interpretation in {\it Our Twenty-First Century Eighteenth-Century War}  comparing with the present war of Russia against Ukraine: {\it The eighteenth-century wars fought by rival European monarchs who could all converse in French with each other, were enviously admired in the bloody twentieth century, because they allowed much commerce and even tourism to persist --- utterly unimaginable even in Napoleon's wars, let alone the two world wars --- and because they ended not in the utter exhaustion of the collapsing empires of 1918, nor in the infernal destructions of 1945, but instead by diplomatic arrangements politely negotiated in-between card games and balls.}
 
 \subsection{The begin of Kirwan's `stellar decade'}

The year 1777, when Kirwan decided to return to London until 1787 again, was  the beginning of his public scientific life, which Sally Newcomb \cite{newcomb2012} called  his `stellar decade'. There he increased his scientific contacts by participation on regular meetings, respectively organizing these himself at his home in Newman-street, 11. There he gave lectures in chemistry for young people of high society \cite{scott1979} and invited friends every Wednesday evening. He regularly took part at the  meetings of the Royal Society of which he was  Fellow since 1780.

 In the same year in December he was one of the initiators of the Chapter Coffee House Society. Since visitors from other British cities and from the continent were also invited to these meetings he could further increase his contacts. Richard Kirwan was  connected as already mentioned to the `Lunar Men' a circle centered in Birmingham. 
He stayed in touch with his scientific friends by writing and receiving letters. In this way  knowledge has been disseminated in Europe and  the United States since the 16th century, while the publications in scholarly journals  became increasingly important. This worked even in the second half of the 18th century despite the current wars between the countries. However, this was not always possible through official channels and colleagues often carried letters across borders and through hostile countries.

So Kirwan began publishing in the Philosophical Transactions of the Royal Society London. He was already an experienced chemist through his studies in his private laboratory and well-informed due to his intense studies of the scientific literature. Thus, it was not surprising that already in 1782 {\it  Richard Kirwan was given the Copley Medal `as a reward for the merits of his labours in the science of chemistry'} \cite{Bektas1992}. His work on chemical affinity, published a year earlier, is also named in this context. 

Apart from this work, three important books were published: in 1784 {\it Elements of Mineralogy} and at the end of his stay in London in 1787 {\it An Estimate of the Temperature at Different Latitudes} and {\it An Essay on Phlogiston and the Constitution of Acids}. His books and journal publications  found in Europe great resonance and were translated into French, Russian,  Spanish, and most prominently into German. Many of his articles appeared in Kirwan's German colleague Lorenz Crell's journal {\it Chemische  Annalen}. This increased Kirwan's  international reputation, especially in Germany. All these activities led to honors  by scientific institutions in several countries during this time. He became honorary member of the academies of Berlin, Dijon, Uppsala, Stockholm (1784), Manchester (1785), Philadelphia (1786), a fellow of the Royal Society of London (1780), and the Mineralogical Society of Jena.

 \subsection{The scientific work of Kirwan  in Lichtenberg's lectures}

 \begin{quote}{\sc  No one has done more merit in this than the Englishman (!) Kirwan, member of the London Society. His book {\it An estimate of the Temperature of different Latitudes} (London 1787), which was also translated into German by K\"uhn, is a major work on this subject} (Lichtenberg \footnote{Niemand hat sich hierin [der Meteorologie] mehr Verdienste erworben, als der Engl\"ander (!) Kirwan, Mitglied der Londner Societ\"at. Sein Buch {\it An estimate of the Temperature of different Latitudes} (London 1787), das durch K\"uhn auch ins Deutsche \"ubersetzt wurde, ist ein Hauptwerk \"uber diesen Gegenstand. (Lichtenberg in seiner Vorlesung `W\"arme der Atmosph\"are' in \cite{Lichtenberg_gamauf}).} in his lecture `Heat of the Atmosphere' in \cite{Lichtenberg_gamauf})  \end{quote}

G\"ottingen was one of the most recognized universities in Germany, in particular, in natural sciences and Lichtenberg  was one of the extraordinary university teachers at that time.  The university was founded in 1737  by George II --- King of Great Britain and Elector of Hannover at that time and since then had a lot of students also from Great Britain. Lichtenberg was in Britain in 1770  for a short time and over one year from August 1774 till the end of December 1775. 

\begin{figure} 
\centering\includegraphics[height=5.5cm]{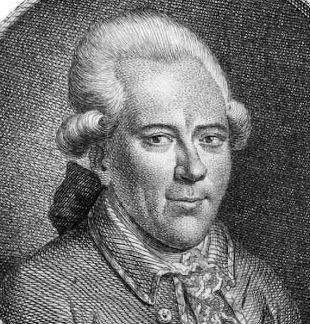}\hspace{0.1cm}\includegraphics[height=5.5cm]{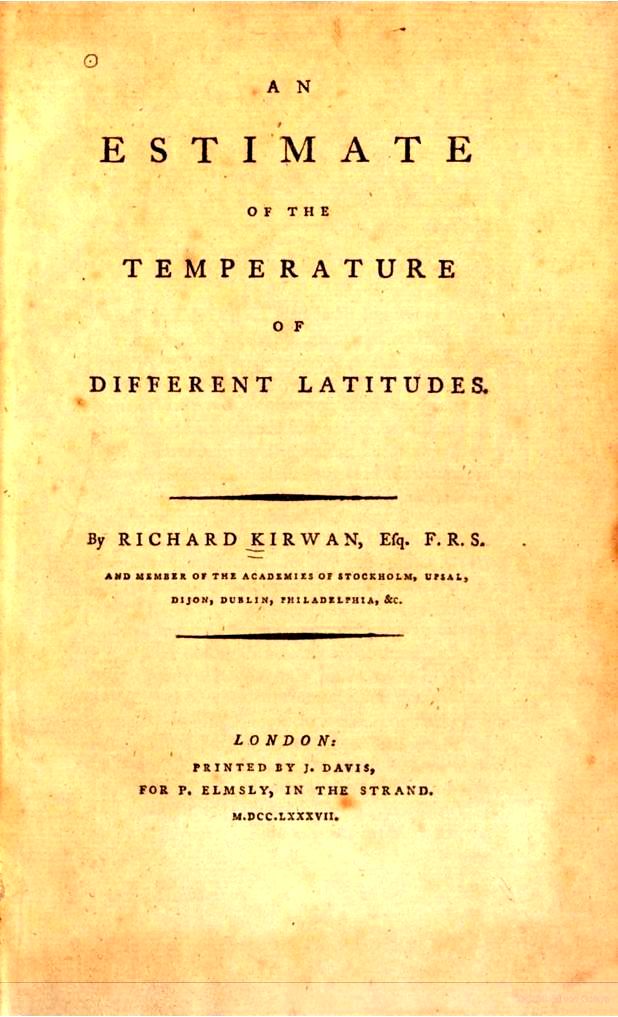}\hspace{0.1cm}\includegraphics[height=5.5cm]{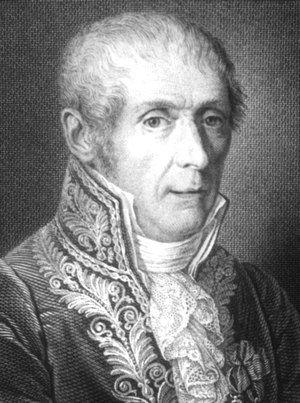}
\caption{(Colour online) G. Ch. Lichtenberg, Kirwan's publication on meteorology \cite{kirwanTemp1787}, A. Volta.}
\end{figure} 

In the year 1777 Lichtenberg's colleague and best pal Johann Christian Polycarp Erxleben (1744--1777) died. Georg Christoph  Lichtenberg (1742--1799) took over his duties as professor for physics at the University of G\"ottingen. In order to support Erxleben's wife he used for his lectures Erxleben's compendium {\it Anfangsgr\"unde der Naturlehre} ({\it Foundations of the Natural Sciences}) \cite{erxleben} extended with his own comments and references. He began his lectures in 1778. The first compendium revised by Lichtenberg, the third edition, appeared in 1784 and the last, the sixth edition, in 1794. In his lectures he tried to be up to date with the latest science\footnote{Only a Polish, Russian and Danish translation  appeared \cite{beaucamp1994}.}. This can be seen in the published notes that he included into the new editions of Erxleben's {\it Naturlehre}, in the lecture notes of his students and in his private notes, which he only used in preparation for his lecture, or simply as brainstorming, for future work or just for fun \cite{tester2016}. 
With these editions one can follow the chemical revolution, the move away from the phlogiston system towards the new French chemistry. The two most prominent representatives of these systems were Richard Kirwan and Antoine Lavoisier.

The fact that Lichtenberg was also well informed about Richard Kirwan's other publications can be seen from the references he added in his lectures. For example, the student Gottlieb Gamauf (1772--1841) reports  in the chapter on meteorology of his lecture notes the citation quoted above (for K\"uhn's translation see \cite{kirwanTemp1788}). 
However, not only Lichtenberg was impressed by Kirwan's work in meteorology but also Alessandro Volta (1745--1827),
 who complained about the unsatisfactory situation of meteorological observations in Italy. Volta's opinion is cited  \cite{Ciardi2001} that this gap {\it in beautiful and learned Italy}, had been particularly revealed in {\it a work that was truly outstanding} by Richard Kirwan referring to the Italian translation of Kirwan's work \cite{kirwanTemp1790}. 

Kirwan wrote to this topic: {\it There is no Science in the whole circle of those attainable by man, which needs the {\sc conspiracy}, if I may so call it, of all nations, to bring it to perfection as Meteorology; nor is there any, perhaps, more conductive to his security and comfort. It is not sufficient that observations should be made in one city, in one kingdom, or even in one hemisphere} \cite{kirwanTemp1787}.  This shows how far ahead of his time Kirwan was. He viewed meteorology as a global scientific task that requires the {\bf conspiracy of all nations}, quite similar as this was already recognized for astronomy (e.g., the Venus transition).  He believed that a careful examination of meteorological records would reveal the spatial and temporal dimensions of weather and climatic change.  The physics Nobel Price 2021 made this vision clear once again. Furthermore, it was made clear that not only global observation but also global responses are necessary. 

Lichtenberg (see the chapter {\it Von der Luft} in \cite{Erxleben1794} ) and Volta \cite{Abbri2000} were of course also interested in the ongoing {\sc chemical revolution} both being more or less supporters of the phlogiston theory with their own notion\footnote{Hasok Chang  called Lichtenberg a fence-sitter \cite{chang2012}).}. Thus, they knew and followed the chemical work of Richard Kirwan and the discussions about the composition of air and the understanding of combustion experiments.  

On  August 6th 1788 Lichtenberg writes in his `Heffte'  (`booklet')\footnote{Kirwan behauptet nun die inflammable, reine Lufft [Wasserstoff] sey dieses Phlogiston, er hat die Meinung w\"urklich vortrefflich unterst\"uzt, zumal in s. neusten Werk \"uber das Phlogiston, und ich glaube auch da{\ss} die Meinung von der w\"urklichen Existentz eines Phlogistons immer hafften wird, wenigstens m\"ussen die Versuche decisiver ausfallen, als bisher --- Allein nun hat sich seit geraumer Zeit unter der Fahne des HE. Lavoisier und de la Place eine Secte entsponnen, die l\"augnet g\"antzlich alle Existentz eines Phlogistons, und man f\"allt ihnen Hauffenweise zu. Die Neuheit und Sch\"onheit der Sache tr\"agt etwas dazu bey. --- Doch mu{\ss} ich sagen, da{\ss} die gro{\ss}en Verdienste dieser M\"anner Respect gebietet. Man mu{\ss} bey solchen Dingen nicht entscheiden wollen, wenn man auch eine Zeitlang hindurch von einigen f\"ur ein Nichtkenner oder wenigstens f\"ur furchtsam gehalten. }
: {\it Kirwan now claims that the inflammable, pure air [hydrogen] is this phlogiston, he has really excellently supported the opinion, especially in his latest work on phlogiston, and I also believe that the opinion of the real existence of a phlogiston will always hold, at least the experiments must be more decisive than before --- But now it has for some time a sect been spun under the flag of the gentlemen Lavoisier and de la Place  that completely denies the existence of a phlogiston, and one falls to them in droves. The novelty and beauty of the thing adds something to it. --- But I must say that the great merits of these men command respect. You don't have to want to decide on such things, even if for a while you were considered by some to be ignorant or at least timid} \cite{lichtenberg_Hefte}.

On his grand tour from 1781 to 1782, Volta traveled to most of Europe's major scientific centers, including the French Academy in Paris, and demonstrated his electrical equipment and inventions to important figures such as Antoine Lavoisier and Benjamin Franklin. Volta became known outside Italy.  Judging by Volta's own testimony, he particularly enjoyed the gatherings in London that brought together instrument makers, independent scholars, and fellows of the Royal Society. When he visited England he had close ties to the club called the Chapter Coffee House Society. He particularly vividly described a meeting at Abraham Bennet's home in which Cavallo, Kirwan, and Walker also took part  \cite{pancaldi2003}. 
  
 \section{Back to Dublin}
 
 \begin{quote}{\sc With labour and money nothing is impossible} (Favourite maxim of Richard Kirwan reported by Lady Morgan in \cite{Morgan1829})
 \end{quote}
 
 After the successful  years in London Kirwan returned for his later life back to Dublin.  His work so far comprised chemical papers published in the Philosophical Transactions of the Royal Society, the different editions of the {\it Essays on Phlogiston}, the {\it Elements of Mineralogy} and the widely recognized meteorological work {\it An Estimate of the temperature of different latitudes}.   Back in Ireland he shifted his work to new fields related to chemistry and useful for the development of his country. 
 
 An excellent example of this type of publications is his book on {\it Manures [Fertilzers]}  \cite{kirwanManures} which was first published\footnote{One of the rare books of Kirwan edited in Dublin.} in 1794, reaching six editions by 1806 and was published  in the United States in 1807. The book is a comprehensive guide to the different types of manures  adjusted to the different types of soils.  The American President Thomas Jefferson (1743--1826) wrote in a letter from 23 March 1798 to William Strickland \cite{jefferson1798}: {\it I am much indebted to you for mr Kirwan's charming treatise on manures. Science never appears so beautiful as when applied to the uses of human life, nor any use of it so engaging as agriculture \& domestic economy. Doctr. Home had formerly applied the doctrines of chemistry to the analysis of soils \& manures, but the revolution in that science had required the work to be done over again, \& gives to mr Kirwan's the entire merit of a new work.}

Two years earlier in 1792 Kirwan  had advocated for the acquisition of the Leskean mineral collection~\cite{golinsky2016,ibler2015}, which was one of the largest in Europe. Kirwan could secure it for himself, the country and his colleagues [e.g., Robert Jameson (1774--1854) from Scotland]. He was already able to use it for the second edition of {\it The Elements of Minerology}. Today, the collection is exhibited in Natural History Museum in Dublin.
 
 With the {\it Geological Essays} \cite{kirwan1799} he  used   his knowledge of chemistry and mineralogy on the one hand. On the other hand, he fought against the newly formulated theory by James Hutton (1726--1797). Hutton, a Scottish geologist was a so-called  Plutonist. He proposed that  rocks  formed from melt, while Neptunists like Kirwan followed the Genesis story that they formed from a solution, the ocean, after Noah's Flood. He opposed Hutton's theory primarily because it contained infinite cycles of geological processes, which he believed contradicted the Bible. Kirwan's geological interest was also driven by improvement of economical interests using Ireland's ore deposits.  In 1800 he was appointed Inspector General of His Majesty's Mines in Ireland. Kirwan's expertise furthered the mining industry in Ireland~\cite{newcomb2011}.
 
Kirwan's work  {\it An estimate of the Temperature of different Latitudes} and its impact was already mentioned. Now back in Ireland, he further promoted the importance of collecting weather data and regularly published measurements he himself took at his home No. 6, Cavendish Road.  In 2007 Charles Mollan in his biography on Richard Kirwan \cite{mollan2007,mollan2007a} wrote: {\it I wonder what his reaction would be if he could come back to view the enormous calculating power of the modern supercomputers used in modern meteorology, and their improving, but still inadequate, ability to predict our weather for more than a short time span? And his comment about our ability to foresee changes is particularly poignant at the time I write this, only a short while after the appalling devastation resulting from Hurricane Katrina in New Orleans and surrounding areas in the United States. The meteorologists gave ample warning, but the reaction at local and national level was totally inadequate.} And today we can say although it {\it is only in a few places in the world, such as at ECMWF\footnote{European Center  for Medium-Range Weather Forecasts.}, that all the different components of observations, prior physical knowledge, and Bayesian learning methods (including data assimilation and machine learning) can be brought together to generate the highest-quality Earth system analyses and forecasts} \cite{geer2023}.  AI makes possible predictions of hundred of weather variables over 10 days at 0.25$^\circ$ resolution globally in under one minute using 40 years of weather data \cite{lam2023}.
   
In addition to all his scientific work, the Unitarian Kirwan maintained social contacts in different societies \cite{Abbas2017}. He was active in the the Royal Irish Academy, where was a committeeman and librarian, and the Catholic physician, William James MacNeven was the Academy's secretary. Together with the Anglican Earl of Charlemont, the President of the Royal Irish Academy at that time, he hosted  literati meetings at his house and played a prominent role in the cultural life.

\subsection{Conversion from Phlogiston 1791}

\begin{quote}
{\sc Although the revolutions in philosophical opinions are of comparatively little importance; yet they often excite some astonishment, especially when they have previously been vigorously defended. For the same reason you may perhaps be a little surprised if I confess to you that I have now given up Stahl's system of phlogiston.}
(Richard Kirwan \footnote{Obgleich die Revolutionen in philosophischen Meinungen verh\"altni{\ss}m\"a{\ss}ig nur von geringer Wichtigkeit sind; so erregen sie doch \"ofters einige Verwunderung, besonders wenn jene vorher mit Lebhaftigkeit vertheidigt worden waren. Aus demselben Grunde m\"ogten Sie sich vieleicht ein wenig verwundern, wenn ich Ihnen gestehe, da{\ss} ich das Stahlsche System vom Phlogiston anjetzt aufgegeben habe.} 1791 \cite{Kirwan1791})
\end{quote}

The analysis of air and its role in combustion experiments in the last half of the 18th century led to the discovery that the `element' air consisted of different gases namely: nitrogen N$_2$  {\sc noxious air}  (Daniel Rutherford 1772), oxygen O$_2$ {\sc dephlogisticated air} (Joseph Priestley 1774, two years after Carl Wilhelm Scheele's {\sc fire air}, but which was published in 1777), carbon dioxide CO$_2$  {\sc fixed air} (Joseph Black, 1756). Already earlier hydrogen H$_2$ {\sc flammable air} (Henry Cavendish, 1722) was found. In 1781 Cavendish was the first person to notice that this gas produces the `element' water H$_2$O when burned. When Antoine Lavoisier and Pierre Laplace replicated Cavendish's discovery in 1783, they named the element hydrogen. The phlogiston system persisted but different identifications of phlogiston were published.

Kirwan was Lavoisier's main rival in the interpretation of the combustion experiment. Therefore, immediately after Kirwan had published his {\it An Essay on Phlogiston} in 1787 \cite{kirwanPhlogiston1787}, the French chemists replied in 1788 with the translation of Kirwan's work  by M.-A. P. Lavoisier including comments of  MM. de Morveau,  Lavoisier, Laplace, Monge, Berthollet, and de Fourcroy \cite{kirwanPhlogiston1788}. Now Kirwan  responded in 1789 in the translation of the French edition back to English \cite{kirwanPhlogiston1789}. 

After all his struggling  and the overhelming arguments of the French chemists, Kirwan abandoned phlogiston theory in 1791. The withdrawal of phlogiston theory by Kirwan (see the quote above) was announced in several places. Most prominently in Germany \cite{Kirwan1791} by a short note  which was sent with a longer paper\footnote{His thought about the different modes of reasoning \cite{kirwan1791b} translated by Carl [!] Crell.} to Lorenz Crell (1744--1816). Crell a proponent of the phlogiston theory popularized Kirwan in Germany by translation of many of Kirwan's work.

Lichtenberg, who followed Kirwan's research, reviews the chemical revolution\footnote{For a more explicit presentation of Lichtenberg's opinion see G. Winthrop-Young thesis 1991 the chapters  {\it Die chemische Revolution [The chemical revolution]} and  {\it Die Omnipr\"asenz der Elektrizit\"at [The omnipresence of electricity]} \cite{Winthrop1991}.} in the introduction to the sixth and last edition  of his lecture textbook {\it Exleben's Naturlehre} \cite{Erxleben1794}. Lichtenberg states there\footnote{da{\ss} die vorz\"uglichsten St\"utzen derselben [der antiphlogistischen Chemie] nicht franz\"osischen Ursprungs sind.}: {\it that the most excellent supporters [of the antiphlogiston chemistry] are not of French origin} and  Winthrop-Young explains that this probably refers to Kirwan and Black `who joined their camp' \cite{Grison1994}. Priestley however, did not change his mind.

In his waste books \cite{lichtenberg_sudelbuch,tester2012} Lichtenberg noted\footnote{Da{\ss} Kirwan \"ubergetreten ist, ist nicht zu verwundern, sein System enthielt schon zu viel von dem anderen, sein Phlogiston hat schon zu viel von dem Hydrogen der anderen, oder ist es schon w\"urklich. [J 1723].}: {\it It is not surprising that Kirwan has converted, his system already contained too much of the other, his phlogiston already has too much of the hydrogen of the others, or is it already real.} The new chemistry could not be stopped although even at that time a lot of  prominent scientists did not change their position to the phlogiston theory. 

Indeed the situation was far from a satisfactory explanation.  Hasok Chang \cite{chang2012} meant: {\it Lavoisier's chemistry never explained why chemical reactions happened, and phlogiston was later seen to have held the conceptual space that chemical potential energy would fill. On the other hand, phlogiston was even at the time commonly identified with electricity, and could easily have been kept and developed into the concept of free electrons.} There is an ongoing discussion between philosophers and historians of science about the assessment of the chemical revolution \cite{kush2015,klein2015,chang2015}, and its interpretation \cite{jacoby2021}.  Continuing the further historical path, we now have reached the state of knowledge represented by the Nobel Prize in Physics 2023 for the study of electron dynamics during chemical reactions \cite{press2023} basis for the important new scientific field of attochemistry \cite{nisoli2019}. In any case, according to Mauskop \cite{mauskop2002}, one can say about Kirwan: {\it If Kirwan was indeed being telling the truth about why he did so --- and there is no reason to doubt his sincerity --- this abandonment of phlogiston for the new chemistry is indeed noteworthy. For it represents a remarkable case where experimental evidence appears to have played the critical role in impelling a senior scientist to abandon the paradigm of which he had become the major expounder, and to accept the new paradigm, which he had only recently so strongly opposed.}

\subsection{Kirwan's thoughts on magnetism}

\begin{quote}
{\sc A {\bf   magnet} therefore is a mass of iron, \dots whose particles are arranged in a direction similar to that of the great internal central magnets of the globe. This I call the {\bf  magnetic arrangement.}} (Kirwan 1797 in \cite{kirwan1797})
\end{quote}

\begin{quote}{\sc Do not use any more the word {\bf theory} in the treatises on {\bf Fire, Electricity},  and {\bf Magnetism}, and on many other subjects in the physics compendium; but {\bf facts}  and {\bf suppositions}; {\bf manner of representing}.} (Lichtenberg\footnote{In dem Compendium der Physik nicht mehr das Wort Theorie zu gebrauchen, bey der Lehre vom {\bf Feuer}, der {\bf Elektricit\"at} und dem {\bf Magnetismus}, und bey vielen anderen; sondern {\bf Facta} und {\bf Muthma{\ss}ungen}; {\bf Vorstellungsart}. \cite{lichtenberg1806} Lichtenberg wanted to write an own physics compendium and he made notes for himself; but the project was never realized.} \cite{lichtenberg1806} before 1789) \end{quote}

The connection between chemistry and electricity gained importance in the last decade of the eighteenth century.
 Other phenomena in nature like aurora borealis  were connected with the earth's magnetic field. Lichtenberg noted already in 1773 in his scrapbook\footnote{Was ist das Nordlicht, die magnetische Materie?} {\it What is the Northern Lights, the magnetic matter?} \cite{lichtenbergC178}.   No connection between electric and magnetic effects were recognized, but both phenomena were in general explained analogous by the behavior of etherlike fluids separating or concentrating in solid matter. The discussions in the various meetings at the Royal Society, in the Chapter  Coffee House Society or the Lunar Society  kept  Richard Kirwan updated with the ongoing speculations about this matter. When Alessandro Volta visited England he had particularly close ties to the club called the Chapter Coffee House Society \cite{pancaldi2003}. For instance,  in  the minutes of the Chapter Coffe House Society \cite{levere2002} in the meeting on February 4th 1785 one reads: {\it Mr. Kirwan is strongly inclined to believe that the more our knowledge of heat and Electricity increases the more evident their connection will appear}, and on March 4th {\it Nicholson read a paper tending to point out those Phenomena of Electricity \& Magnetism that resemble each other.} Then, on December 8th 1786, an experiment is reported {\it made in London (it was observed) by Mr. Volta, who supposed the electricity to be negative} and it is reported that iron {\it when heated {\bf red hot} is found no longer magnetic, loosing then its attractive power.}  
  
There were also  other ideas on magnetic solids not related to an etheric fluid but to small `atomic' magnets, that constitute the magnets especially iron. Mathias Gablers  \cite{gabler1781} suggested this in 1781 and David Rittenhouse' \cite{rittenhouse1786} in 1786. They also speculated that these elementary magnets are rotatable in order to be arranged by the external field of the earth.   Lichtenberg treated the topic magnetism in his physics lectures \cite{mautner_miller}. They are documented by Gamauf \cite{Lichtenberg_gamauf}, who heard these lectures in the years 1793--1796 and published his script in the years 1811--1818. Gamauf \cite{Lichtenberg_gamauf} cited his teacher Lichtenberg\footnote{Und da hat denn Gabler folgenden sch\"onen Gedanken: Das Eisen bestehe aus lauter atomischen Magneten, deren jeder so gut seine Polarit\"at hat, wie man sie an der Nadel bemerkt; allein sie liegen alle unregelm\"assig durcheinander, da{\ss} sie keine magnetische Erscheinung \"au{\ss}ern k\"onnen.... Bringt man hingegen Eisen einem Magneten nahe, so kommen die kleine Magnete alle in Ordnung, alle S\"udpole z.B. gegen den Norpol des Magneten; bringt man  ihn wieder hinweg so \"uberwindet die Koh\"asionskraft des Eisens die magnetische Kraft des Eisens die magnetische Kraft der kleinen Atome, und sie kommen wieder in Unordnung.}: {\it  And then Gabler\footnote{Gabler's theory \cite{gabler1781} from 1781 is rarely mentioned, but see \cite{wiederkehr}, on the other hand, neither Rittenhouse nor Kirwan is mentioned.} has the following nice thought: Iron consists of nothing but atomic magnets, each of which has its polarity as you can see it on the needle; but they are all mixed up irregularly so that they cannot lead to any magnetic phenomenon.... If, on the other hand, you bring iron close to a magnet, the small magnets all come into order, all south poles, e.g., against the north pole of the magnet; if you take it away again, the cohesive force of the iron overcomes the magnetic force of the iron, the magnetic force of the small atoms, and they become disordered again.} And he noted in his `B\"uchelgen' for his lecture\footnote{Abends den 26. M\"artz (1798?): ... Gabler's Theorie Rittenhouse hat dieselbe Lehre ... Elementarische Magnetnadeln... Also den Elementarischen Magnetnadeln die Richtung geben.}  and in his marginal notes to the textbook of Erxleben\footnote{Gablers Theorie tr\"agt auch D. Rittenhouse in den Trans. of the American society vor. (see \cite{rittenhouse1786}).}: {\it Gabler's theory is also presented by D. Rittenhouse in the Trans. of the American society}. 

However, these ideas seemed to be not so popular until Richard Kirwan published his independently formulated {\it Thoughts on Magnetism} in the year 1797 \cite{kirwan1797}. He also supposed `atomic' magnets but in addition he postulated: {\it In consequence then of the universal law of attraction of the particles of matter to each other, these internal magnets exert a {\it double} power of attraction} analogous to the power of {\sc crystallization}. {\it By crystalization I understand that power by which the integrant particles of any solid possessing sufficient liberty of motion unite to each other, not indiscriminately and confusedly, but according to a {\bf peculiar uniform arrangement}}[pronunciation by the author], {\it so as to exhibit in its last and most perfect age stage regular and determinate form.} And he concluded with the quotation given at the beginning of this section. This {\bf magnetic arrangement}
is a surprising early statement of an emerging order resulting due to an interaction in a manybody system as we would state today. He even discriminates between the forces leading to the crystalline order and the magnetic order. {\it For in this case it exerts a double attractive power, that of the particles of iron to each other, \ldots,  and that of crystalizing bodies, which we have also seen to be indefinitely great.}

The {\it Thoughts on Magnetism} were translated into German, first in 1800 \cite{kirwanG1800} by Ludwig Wilhelm Gilbert (1769--1824; the editor of the {\it Annalen der Physik}) and a  second time in 1801 by Lorenz von Crell (1744--1816) in the fifth volume of the translation of Kirwan's {\it Physisch-chemische Schriften [physical-chemical Writings]} \cite{kirwan1801crell}.  Crell writes\footnote{Er [Kirwan] vergleicht hier, auf eine scharfsinnige, vor ihm noch nicht angewandte, Art die Erscheinungen der Crystallisation im Allgemeinen, und ihre Gesetze, mit den Erscheinungen der Anziehung des Eisens vom Magneten. Er glaubt eine gro{ss}e Aehnlichkeit zwischen denselben zu finden, und sie daher aus einer allgemeinen Ursache ableiten zu k\"onnen.}to the {\it Thoughts}: {\it Here he [Kirwan] compares the phenomena of crystallization in general, and its laws, with the phenomenon of the attraction of iron to the magnet, in a way that has not yet been used before him. He believes that he can find a great similarity between them and can therefore derive it from a general cause.}

Since then Kirwan's {\it Thoughts} were taken up  under the name of the Rotation Hypothesis in contrast to the Separation Hypothesis of the fluid model  in Encyclopedia articles on magnetism \cite{fischer1805,fischer1808,lamont1867,auerbachHandbuch} in Germany. Later they were also mentioned in 1883 in the Encyclopedia Britannica \cite{chrystalEB1883} through the article by George Crystal (1851--1913). Most prominently Ernst Ising started the historical path of knowledge, to the model he formulated, with Kirwan's {\it Thoughts}. But this reference was not included in Ising's publication in {\it Zeitschrift f\"ur Physik} and Kirwan's {\it Thoughts} dropped out of the history of magnetism although this time\footnote{William B. Ashworth judged when discussing Kirwan's chemical and geological contributions: {\it He had the dubious distinction of being on the wrong side in both controversies, yet he was the one that the proponents of Lavoisier, and of Hutton, feared the most, and spent the greatest time trying to refute }\cite{ashworth2023}.} Kirwan was on the `right side'. Exceptions of citing Kirwan's {\it Thoughts} in this context are the papers by Kiozumi \cite{koizumi1983} and quite recently by Coey and Mazaleyrat \cite{coey2023}.
 
Let this section close with how Lichtenberg  read the book of nature in comparison with Kirwan (see the quote in section~\ref{late}).  Lichtenberg, quite similar, meant\footnote{Vor Gott ist nur Eine Naturwissenschaft, der Mensch macht daraus isolirte Capitel und mu{\ss} sie, nach seiner Eingeschr\"anktheit machen. So lange als die Capitel nicht zusammen passen wollen, liegt irgendwo ein Fehler, in den einzelnen besonders, oder in allen.} \cite{Erxleben1794} in 1794: 
\begin{quote}{\sc Before God there is only one natural science; man makes isolated chapters out of it and has to make them according to his limitations. As long as the chapters don't fit together, there is a mistake somewhere, especially in some of them, or in all of them.} \end{quote}

 Both seem to recognize the emergence of different disciplines, and they also recognized that there is a common basis. Both Kirwan and Lichtenberg advocated the importance of experimental investigations but also sensed a common underlying level connecting the different natural phenomena. The discoveries of the nineteenth century on the path of knowledge led to this underlying level and showed  how the chapters could be read.

\section{The Irish Rebellion and the Act of Union}

\begin{quote}{\sc Perhaps the share which some noted scientific men have lately taken in the convulsions of a neighbouring country [France], may seem to invalidate the above assertion [The enterprize exploring the Constitution of nature for lasting splendour on their reigns], but it should be remarked, that of the votaries to {\bf natural knowledge}, many became the victims of that direful tyranny, suffering either death as Lavoisier and Died[t]rich, or exile as Bournon, De Mazieux, La Peyrouse \&cc. The few whose names still remain enrolled in the ever execrable annals of anarchy, were nevertheless in reality guiltless of its enormities, being restrained from opposition, by the then all prevailing terror. The pretended philosophic reformers of metaphysics, morality, and politics, and the frantic enemies of christianity alone prepared, excited, and acted those attrocious tragedies, in comparison of which the accumulated cruelties of ancient tyrants, and of pagan and christian persecutions, are lost to the sight. } (Richard Kirwan in the preface to the {\it Geological Essays} 1799 \cite{KirwanGeolog1799})\end{quote}

Richard Kirwan usually took no part in public politics \cite{McLaughlin1940IV} although he had friends who strongly supported the political movements for republic constitutions and freedom --- the most prominently was Priestley in Britain and many other in France like Guyton de Morveau. The situation changed in the years after the French Revolution. 
Politics had begun to dominate over religion. Paine's Rights of Man had become the Bible of Belfast, and the Orange Order were founded in 1795 by Protestants in Ireland who feared the growing Catholic influence and political involvement on the island. Hely Hutchinson (1724--1794) provost of Trinity College admitted Catholic students to the college a decade before the legal relaxation of the penal code in 1793, and he spoke in the Irish Parliament in favor of establishing a Chair of Catholic Theology in the University \cite{McLaughlin1940III}.

Already on October 1791  the {\sc Catholic Society} was founded in Dublin and a declaration\footnote{{\it In  the present enlightened and improving period of society, it is not for the {\sc Irish Roman Catholics} alone to continue silent. Not accused of any crime; not conscious of any delinquency, they suffer a privation of rights and conveniencies, the penalty reserved in wise states for offences of atrocious magnitude} \cite{McKenna1794}.}  was issued, written by the physician McKenna (1765--1808), calling for the abolition of the entire penal code. The language and arguments of this declaration, which advocated {\it a spirit of harmony, and sentiments of affection} between Irishmen, bear a striking resemblance to the first public pronouncements of the {\sc Society of United Irishmen}, which was founded in Belfast the same month. A {\sc Dublin Society} was then established in November. The primary objective of the {\sc United Irishmen} was parliamentary reform, and they drew much of their inspiration from the French Revolution, but the society was also a response to the political ferment and possibilities opened up by the catholic question \cite{Smyth1998}. In November The Dublin Society of United Irishmen was formed.

There is a debate as to whether  Richard Kirwan was a member of the United Irishmen. In his thesis H. Abbas gives a list of the Dublin Library Society Founding Members who were also members in the Dublin Society of United Irishmen (see table 19 in \cite{Abbas2017}) and Richard Kirwan is included.  It is said that he had joined the society in 1795. However, the list given by McDowell  mentions Martin Kirwan \cite{McDowell1940}. But Dixon said Kirwan was sworn into the United Irishmen by MacNeven in 1795/96 \cite{Dixon1971}.  In any case there is no evidence that Richard Kirwan ever attended meetings. Nevertheless, he rendered useful service in 1798 when he noticed that raids were carried out in Abbebey Street (Dublin). He immediately advised  his friend Sampson to burn compromising papers before the raiders reached the house  \cite{McLaughlin1940III}.

The mainspring of Kirwan's activity was the desire to apply the fruits of science for the benefit of his country without neglecting pure science. He had refused the offer\footnote{{\it In the hope that his great influence would help in promoting the Union} \cite{reilly1930}. } of a baronetcy by Lord Castlereagh (1769--1822) when the latter had asked him to exert his great influence in favour of the Union \cite{McLaughlin1940IV}.  Castlereagh worked to suppress the Rebellion of 1798 and to secure passage of the Irish Act of Union  in 1800 and was the Chief Secretary for Ireland at the time of the Union. He is named\footnote{A citation I take from Ralph Kenna's talk {\it From the Ising model to a distant academic world and back} given in July 2019 in Coventry \cite{ralph2019}.} in P. B. Shelley's {\it The Masc of Anarchy: Written on the Occasion of the Massacre at Manchester (1819)}: {\sc I met Murder on the way --- He had a mask like Castlereagh --- Very smooth he looked, yet grim; Seven blood-hounds followed him.}  

\section{Final projects}

\subsection{A new academy for Dublin}

\begin{figure}[h]
\centering\includegraphics[height=6.5cm]{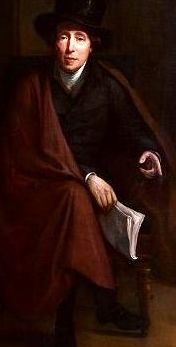}\hspace{0.15cm}\includegraphics[height=6.5cm]{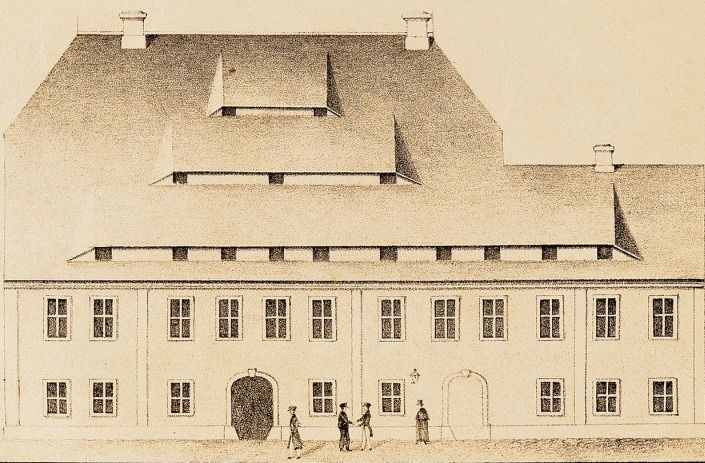}
\caption{(Colour online) (a) Richard Kirwan detail of the painting by Hugh D. Hamilton (1816) and (b) Drawing of the Royal Saxon Mining Academy building (1831) \copyright Media center of TU BA Freiberg. \label{final} }
\end{figure}

The situation of Ireland's mining industry is described by Sally Newcomb \cite{newcomb2011}: {\it Due possibly to its turbulent history, mining in Ireland had not been as pursued until somewhat after that in Germany, Sweden, France, Hungary, Italy, Spain and its possessions, and England. There was no formal structure for studying geology or exploring or regulating mining until well after the establishment of schools of mining in many of those other countries.} Kirwan was in 1800 appointed Inspector General of His Majesty's Mines in Ireland.   Kirwan's expertise should advance the mining industry in Ireland.

To improve this situation, in 1798 he planned to found a mining academy in Ireland similar to that in Freiberg, Saxony, Germany. For that purpose, George Mitchell (1766--1803) and later  others enrolled  as a student at the Mining Academy in Freiberg \cite{ibler2014}. Mitchell invited Friedrich Mohs (1773--1839) to work with them in Freiberg on the planing this academy and invited him to Ireland to discuss the new institute in Dublin under the leadership of Richard Kirwan. Mohs, of course, accepted this honorable invitation which also corresponded to his inclinations.  So he returned to Freiberg from Harzgerode in Saxony-Anhalt and started the preparations for the project with Michell and Robert Jameson (1774--1854) who had meanwhile arrived from Edinburgh. For this purpose they collected books, drawings, models, tools which could be used for teaching at the new institution. The equipment for various devices should then be handed over to the teaching institution. 

As useful as the planned institute in Dublin could have been for Great Britain and Ireland, the political conditions at the time and generally the circumstances delayed its implementation. Moreover, Mitchell's illness and his early death in 1803 and later after Kirwan's death,  the project was never realized. Instead, Mohs moved to Austria in 1802 and became a professor at the Joanneum Graz in 1812. There he developed his famous scale of mineral hardness published in 1822.  Sally Newcomb remarks \cite{newcomb1990} that Kirwan {\it used a comparative, numerical, hardness scale nearly 30 years before Mohs did.} Later Mohs became professor for mineralogy in Freiberg and in 1826  full professor of mineralogy at the University of Vienna \cite{fuchs1843,authier2013}.

\subsection{Final thoughts and honor}
 
In his last years, Kirwan became increasingly interested in fundamental philosophical considerations. He published the {\it Principles and Different Modes of Reasoning} \cite{kirwan_logic1807} and the {\it Principle and Fundamental Object of Science} \cite{KirwanMetaphys1809}. Parts of these where already published in a German translation by Crell \cite{reasoning1791}.  Rick Kennedy discusses Richard Kirwan's {\it Logick}  as an example of an English textbook with extensive interest in reducing good sense to calculus and relates. This eighteenth-century desire to create a calculus for testimony was largely founded on the work of Jakob and Nikolaus Bernoulli.  Kennedy judges {\it Kirwan is the most extensive student in a tradition begun at Port-Royal in the seventeenth century that continued into the early twentieth century} \cite{kennedy2004}. The most interesting point is  the introduction of the mathematics of probability for testimony.

As already mentioned, Kirwan emphasizes that the knowledge about nature is based on precise and careful experiments and on observations, and he cites Newton claiming that same effects should have same causes. He suggests analogy as a heuristic tool for understanding open problems. This he most clearly used  in his {\it Thoughts on Magnetism}. Interesting is his statement where he cites Priestley: {\it The mere difficulty of seeing how a thing can be, must not be an argument against a correct inference drawn from true premises.} He also states: {\it So the probability or truth of any doctrine, argument, or speculative opinion is derived from its conformity or analogy to some other doctrine of whose truth we are certain}, and some pages later: {\it In experimental sciences, they are summaries, or final results, from numerous facts, and are highly useful, as from them, several new facts may often be deduced by analogy.}

In his {\it Metaphysical Essays} \cite{kirwan1806}, he presents three parts: {\it  My aim in the first is to point out the true signification of terms that most frequently occur in treating metaphysical subjects, and explain the nature of objects indicated by some others. \ldots  In the second I have endeavoured to do away with some false opinions respecting the human souls, and particularly its materiality; \dots The existence of the Supreme Being and his attributes, as far as they are discoverable by human reason, form the subjects of the third essay.} His explanations were critically commented 1810 in {\it The Critical Review: or, Annals of Literature} but the final judgment was that {\it these Essays may serve as a good introduction to the study of metaphysics.} 
In summary, while Hume, Locke, and Berkeley all made important contributions to empiricism and philosophy, they held differing views on the nature of reality, the nature of human knowledge, and the extent of skepticism applied to various philosophical concepts. Locke emphasized the role of experience in knowledge, Berkeley denied the existence of a material world, and Hume raised profound questions about causation and the limits of human reason.

 In the year 1808 Richard Kirwan\footnote{Together with Scottish inventor and engineer  James Watt and English physician Edward Jenner.} was elected foreign associates of the French Academy of Sciences (then known as {\it the Institut National})  although the Napoleon war was going on. As already quoted Lavoisier~\cite{Lavoisier1793} had in 1793 remarked {\it the sciences were never at war}, which can be considered  an oversimplification and lead to a ongoing discussion (see, e.g., \cite{beer1960,Lipkowitz2009}). Lipkowitz {\it  argues that wartime politics and the increasingly systematic use of science in support French and British national interests created a rupture in the practices and activities of the European scientific Republic of Letters, a disjuncture that paved the way for the emergence of very different regimes of transnational science in the nineteenth century.} 
 
In Germany, e.g., the growing opposition to the Enlightenment even before the French Revolution was initially fed by fear of Jacobinism, then by resistance to Napoleonic imperialism. Young intellectuals who could not abide the cant and compromises of enlightened absolutism, joined conservatives in denouncing the Enlightenment advocates \cite{hufbauer1982}.  Lady Morgan clearly states Kirwan's opinion: {\it Abhorring the atrocities of the fatal re-action, which retarded the benefits and stained the cause of the French Revolution, he was frank and loud in his reprobation of that ruinous continental war by which the British empire was drained and demoralised, to re-vive pernicious institutions, and restore a race, the antitypes of the unfortunate family which England had herself spurned and dethroned. It was curious to hear him calculate the expenses of this war, and the disbursement which would have been required to build a causeway or pier that should extend across the channel} \cite{Morgan1829}.

\section{Conclusion}

Richard Kirwan made important contributions to the development and constitution of chemistry, mineralogy,
meteorology, geology and the special field in physics of ferromagnetism. All this happened in the second half of his life and was based on a deep enthusiasm for reading books and and making own experiments to understand nature. He created for himself a big library and a network of scientific friends in Europe. And he was convinced on the usefulness of science for the economic and humanistic development of his country. Richard Kirwan could have chosen Ralph Kenna's vision statement \cite{ralphvision}:
{\sc The single motivation for all of my research is curiosity. I constantly hope to achieve greater understanding of various complex systems, from the physical sciences to the sociological and the humanities.}

\section*{Acknowledgements}

It was in 2017 during the Ising Lecture when the decision was made to perform a full English translation of Ernst Ising's  entire dissertation, shortly after we --- Ralph Kenna, Yurij Holovatch, Bertrand Berche and I  --- had completed  a publication on the life and work of Ernst Ising with Tom Ising. I already had a copy of the dissertation and  to our surprise in the introduction there was found the name Richard Kirwan, an Irish colleague of Ralph from the 18th century so to say. Since then I have tried in close contact with Ralph to uncover the story of Kirwan. It turned out that I could always ask my Irish friend Ralph to comprehend a scientist of that time and in the appropriate context. Most important for me was his perspective on the Irish identity of Richard Kirwan, who worked in England and was connected to his European friends. Ralph's focus on Irish culture, his ability to bridge the various fields of  science and humanities was a key driving force crucial for me to continue working on the history of Richard Kirwan.

\newpage

\ukrainianpart

\title{Рiчард Кiрван, ірландський науковець в Європі}
\author{Р. Фольк}
\address{Інститут теоретичної фіизики, Університет Йоганна Кеплера, 4040 Лінц, Австрія}

\makeukrtitle

\begin{abstract}
	Кінець вісімнадцятого та початок дев'ятнадцятого століть довгий час вважалися періодом формування сучасних ірландських політичних
	традицій, таких, як націоналізм, республіканізм і юніонізм. Для Європи це був час змін в науці, переходу від спостережень до експерименту,
	і від припущень до фактів. Річард Кірван був відомим у Європі натурфілософом і шанованою людиною науки свого часу. Навіть під час усіх воєн, 
	його зв’язки з колегами утворювали свого роду мережу, яка вкривала Європу та сягала навіть до Америки. Ця стаття 
	показує на основі декількох прикладів, як ця мережа працювала в час, що відзначався політичними конфліктами та революційними подіями як у науці, так і в суспільному житті.

	\keywords історія науки, історія фізики, природознавство
	
\end{abstract}

\lastpage
\end{document}